\documentclass[conference]{IEEEtran}
\IEEEoverridecommandlockouts
\usepackage{cite}
\usepackage{amsmath,amssymb,amsfonts}
\usepackage{mathtools}
\usepackage{algorithm}
\usepackage{algpseudocode}
\usepackage{graphicx}
\usepackage{subcaption}
\usepackage{textcomp}
\usepackage{xcolor}
\usepackage{enumitem}
\usepackage{tabularx}
\usepackage{adjustbox}
\usepackage{makecell}
\usepackage{titlesec}
\usepackage[a4paper, total={184mm,239mm}]{geometry}

\def\BibTeX{{\rm B\kern-.05em{\sc i\kern-.025em b}\kern-.08em
    T\kern-.1667em\lower.7ex\hbox{E}\kern-.125emX}}

\titlespacing*{\section}
  {0pt}{6pt plus 2pt minus 2pt}{6pt plus 2pt minus 2pt}
\titlespacing*{\subsection}
  {0pt}{4pt plus 2pt minus 2pt}{4pt plus 2pt minus 2pt}

\begin{document}

\title{Area-Efficient In-Memory Computing for Mixture-of-Experts via Multiplexing and Caching \vspace{-10pt}
}

\author{
\IEEEauthorblockN{Hanyuan Gao, Xiaoxuan Yang}
\IEEEauthorblockA{
\textit{Department of Electrical and Computer Engineering, University of Virginia, Charlottesville, VA, USA} \\
\{zqv4ds, xiaoxuan\}@virginia.edu}
\vspace{-3em}
}

\maketitle

\begin{abstract}
Mixture-of-Experts (MoE) layers activate a subset of model weights, dubbed experts, to improve model performance.
MoE is particularly promising for deployment on process-in-memory (PIM) architectures, because PIM can naturally fit experts separately and provide great benefits for energy efficiency.
However, PIM chips often suffer from large area overhead, especially in the peripheral circuits. In this paper, we propose an area-efficient in-memory computing architecture for MoE transformers.
First, to reduce area, we propose a crossbar-level multiplexing strategy that exploits MoE sparsity: experts are deployed on crossbars and multiple crossbars share the same peripheral circuits.
Second, we propose expert grouping and group-wise scheduling methods to alleviate the load imbalance and contention overhead caused by sharing.
In addition, to address the problem that the expert choice router requires access to all hidden states during generation, we propose a gate-output (GO) cache to store necessary results and bypass expensive additional computation.
Experiments show that our approaches improve the area efficiency of the MoE part by up to 2.2x compared to a SOTA architecture.
During generation, the cache improves performance and energy efficiency by 4.2x and 10.1x, respectively, compared to the baseline when generating 8 tokens. The total performance density achieves 15.6 $GOPS / W / mm^2$.
The code is open source at https://github.com/superstarghy/MoEwithPIM.
\end{abstract}


\section{Introduction}


The Mixture-of-Experts (MoE) layer consists of multiple independent modules, i.e., experts, and a gate network that routes inputs to different experts \cite{shazeer2017outrageously}.  During inference, each input activates only a subset of experts. This partial activation approach allows models to expand model capacity while keeping computational complexity manageable. In Transformers, MoE typically replaces feedforward layers and have stood out in vision \cite{fan2022m3vit} and language tasks \cite{jiang2024mixtral, zhu2024llama}.

Process-in-memory (PIM) offers opportunities to accelerate neural network models and attention-based models \cite{yang2020retransformer, jain2022heterogeneous}, and is a promising platform for accelerating MoE \cite{buchel2025efficient}.
However, PIM suffers from the large area overhead of peripheral circuits, especially analog-to-digital converters (ADCs) \cite{andrulis2023raella}.
Furthermore, MoE layers generally suffer from load imbalance, with certain experts receiving more or fewer activations than others \cite{shazeer2017outrageously}.
When applying PIM to accelerate MoE, the impact of imbalance is amplified because the overall efficiency depends on the bottleneck expert in each layer. 
Although many works improve from the software perspective, such as introducing load-related losses during training or limiting the load capacity of experts \cite{shazeer2017outrageously, bengio2015conditional, swichtransformer, lepikhin2020gshard}, they either do not guarantee balanced loads or do so at the cost of model performance.


In this paper, we design a PIM-based architecture for MoE transformers focusing on area efficiency, workload balance, and generation performance. First, MoE inference is inherently sparse, meaning that not all experts are activated simultaneously, providing opportunities to reuse circuit components. Accordingly, we propose a component-sharing approach that reduces the PIM chip area by exploiting sparse expert activation. Specifically, we group the MoE experts and allow several PIM crossbars share the same peripherals based on these groups.

Furthermore, the grouping helps with mitigate workload imbalance. To this end, we design a static load-aware expert grouping scheme and further implement a dynamic scheduling algorithm during the prefill stage to exploit data reuse.

In addition, the auto-regressive nature of large language models limits the efficient deployment of MoE.
For instance, under expert-choice routing \cite{zhou2022expertchoicerouting}, each expert selects multiple tokens, resulting in all inputs retained at every decoding step, which degrades performance severely.
Prior work simplifies the gate to a conventional token-choice \cite{rajbhandari2022deepspeed} design during generation, but the model structure is different from that of training.
To improve generation efficiency, we design a gate-output (GO) cache. The GO cache buffers the gate scores and necessary outputs of the MoE layer, trading storage for performance.
The GO cache works with the key-value (KV) cache to further accelerate the generation process.

The experimental results show that our methods achieve significant improvements. The area efficiency improves by 2.2x compare to a baseline. The latency and energy cost during the generation improve by 4.2x and 10.1x, respectively, compared to 3DCIM \cite{buchel2025efficient}, a SOTA PIM architecture. 
We make the following contributions:
\begin{itemize}[leftmargin=0.8em, labelsep=0.4em]
    \item We propose a component-sharing PIM architecture for MoE transformers, which leverages the inherent sparsity of the MoE layer, implements crossbar-level multiplexing, and effectively reduces the area cost.
    \item We propose the GO cache for expert choice routing MoE, addressing the  generation inefficiency.  The cache works with KV cache to accelerate the auto-regressive process.
    \item We implement the static expert grouping method to balance the overall workload and dynamic fine-grained scheduling to exploit data reuse.
    \item We characterize the MoE with PIM in an operator-accurate simulator with GO cache and scheduling integration.
\end{itemize}

\section{Background and Motivation}\label{sec:background}


\subsection{Mixture-of-Experts}

The Mixture-of-Experts module \cite{shazeer2017outrageously, swichtransformer, lepikhin2020gshard, zhu2024llama, zhou2022expertchoicerouting} consists of multiple experts and a gating network. The gate receives input representations and generates scores between tokens and experts.
In token-choice routing, each token selects experts based on the scores \cite{shazeer2017outrageously}. In contrast, the experts select tokens following the score ranking in expert choice routing \cite{zhou2022expertchoicerouting}. The selected experts and tokens then perform linear operations and accumulation.
The following equations demonstrate the token choice routing MoE \cite{shazeer2017outrageously}, where each token selects top-k experts.

{\small
\setlength{\abovedisplayskip}{3pt} 
\setlength{\belowdisplayskip}{3pt}
\setlength{\abovedisplayshortskip}{2pt}
\setlength{\belowdisplayshortskip}{2pt}

\vspace{-5pt}
\begin{equation}
  G(x) = \mathrm{softmax}(\mathrm{KeepTopK}(xW_G, k)) \label{eq:gate}
\end{equation}
\begin{equation}
\mathrm{KeepTopK}(v,k)_i =
\begin{cases}
v_i, & \text{if } v_i \in \mathrm{topk}(v)\\
-\infty, & \text{otherwise}
\end{cases}
\end{equation}
\begin{equation}
  y = \sum_{i=1}^n G(x)_i E_i(x) \label{eq:calculator}
\end{equation}
}

\noindent In the equations, $G(\cdot)$ and  $E_i(\cdot)$ denote the output of the gate network and the output of the $i$-th expert out of $n$ experts.

Conventional gate network suffers from expert collapse, where only a few experts are activated, leading to load imbalance \cite{shazeer2017outrageously, zhou2022expertchoicerouting}.
To mitigate this issue, prior work proposes loss functions \cite{shazeer2017outrageously} to instruct the training, and expert capacity \cite{lepikhin2020gshard, swichtransformer} to constrain the number of tokens an expert receives.
However, the losses do not provide strict guarantees, while expert capacity strictly restricts the load at the cost of model degradation or reduced flexibility \cite{swichtransformer}.

Beyond these constraints, expert-choice routing \cite{zhou2022expertchoicerouting} adopts an alternative gate design that enables experts to select the top-k tokens.
Expert-choice routing is naturally balanced because each expert selects the same number of tokens.
However, the design is inefficient during autoregressive generation because it requires retaining hidden states of all tokens. Existing methods revert the expert-choice routing to a conventional token-choice routing at inference time \cite{dai2024deepseekmoe, rajbhandari2022deepspeed}, which introduces a mismatch between training and generation.
Therefore, the challenge of supporting expert choice routing with efficient hardware design still remains in generation tasks.

\subsection{Process-in-Memory for MoE}
Process-in-memory architectures reduce the transfer of weight data, achieving high energy efficiency. They are therefore promising for accelerating MoE because MoE consists of many separate linear experts.
Although the MoE model reduces the computational complexity, it still requires a large number of parameters. PIM is well-suited to such workloads with heavy parameters. Several PIM accelerators have been proposed to optimize MoE inference, such as a heterogeneous array \cite{jain2022heterogeneous} and a 3D stack design \cite{buchel2025efficient}. However, these designs focus on mapping a model to the hardware and overlook the opportunities from expert sparse activation and scheduling optimization.

\subsection{Motivation}

When accelerating MoE with PIM, not all PIM crossbars are activated, considering the sparse activation of experts. However, each crossbar requires peripheral I/O. This observation motivates crossbar-level multiplexing in PIM chips. In addition, the peripherals often dominate the area, for example, ADCs account for more than 60\% of the chip area \cite{andrulis2023raella}.
On the other hand, the multiplexing strategy introduces structural contention, as multiple crossbars may compete for shared peripherals. Therefore, a schedule policy is required to fit the peripheral demands across crossbars.
 
In addition to sharing, we must determine which experts are grouped and deployed on crossbars that share peripherals. To reduce contention and performance degradation, we propose a load-aware expert grouping strategy. 
Nevertheless, the sharing and grouping method inevitably exhibits imbalance during inference: the token-expert choices are calculated on the fly, making expert workloads difficult to predict, especially during the prefilling stage with many tokens. 
To address this, we employ a dynamic scheduling policy to enhance data reuse and reduce latency and energy consumption.

Our design targets flexibility to support either token-choice or expert-choice.
For expert-choice, although the load is balanced, auto-regressive generation is not feasible. The challenge is that the gate requires all previous tokens and the current token as input, introducing extra computation and communication.
To overcome this inefficiency problem, we design a gate-output (GO) cache for the MoE layer. The necessary gate results and outputs are cached. Following that, only one newly generated token is processed at each step.

\section{Methods}\label{sec:method}

\subsection{Proposed Architecture}\label{sec:arch}

\begin{figure}[t]
\centering
\includegraphics[width=0.9\columnwidth]{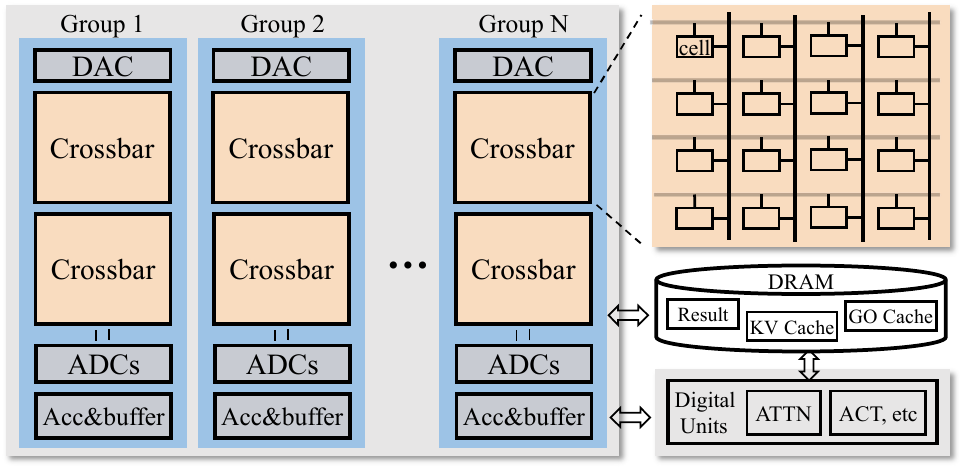}
\caption{Overview of the MoE accelerator with PIM. }
\label{fig:arch}
\end{figure}

We aim to accelerate MoE-based transformers, especially LLMs.
Typically, a transformer block includes a multi-head attention (MHA) layer followed by an MoE module.
We accelerate the linear operations with PIM, because they dominate the overall computation in transformers \cite{rajbhandari2022deepspeed}.
We leave MHA computation to specific digital units, as in \cite{buchel2025efficient}, since attention requires two dynamically generated inputs.
At the architectural level, we enable peripheral sharing across selected PIM arrays through our grouping strategy, as shown in Fig.~\ref{fig:arch}. This strategy eliminates and reuses many PIM components and substantially reduces area overhead.
On the other hand, the strategy may result in contention, but the sparsity of MoE inherently limits the number of potential structural conflicts. This property makes MoE a natural fit for our sharing-based PIM architecture.

\subsection{Expert Grouping for Workload Balance}

To fully exploit the benefits of peripheral sharing, we develop a grouping strategy at deployment time.
Due to the load imbalance in both prefilling and decoding, the grouping directly affects the structural contention among the available resources.
To study the impact of grouping, we develop two heuristics: uniform grouping and workload-sorted grouping. Uniform grouping assigns experts to groups uniformly at random. Workload-sorted grouping indicates that the experts are sorted by their workload and grouped according to their ranking. Specifically, the workload is traced from small samples of datasets.
Then, for a group size of two, experts with the lowest loads and experts with the highest loads will be grouped.
All of these processes are completed before deployment.
After grouping, the overall workload of each group will be statistically similar, thereby successfully balancing load across groups without imposing explicit constraints on experts.

\begin{figure}[t]
\centering
\includegraphics[width=0.96\columnwidth]{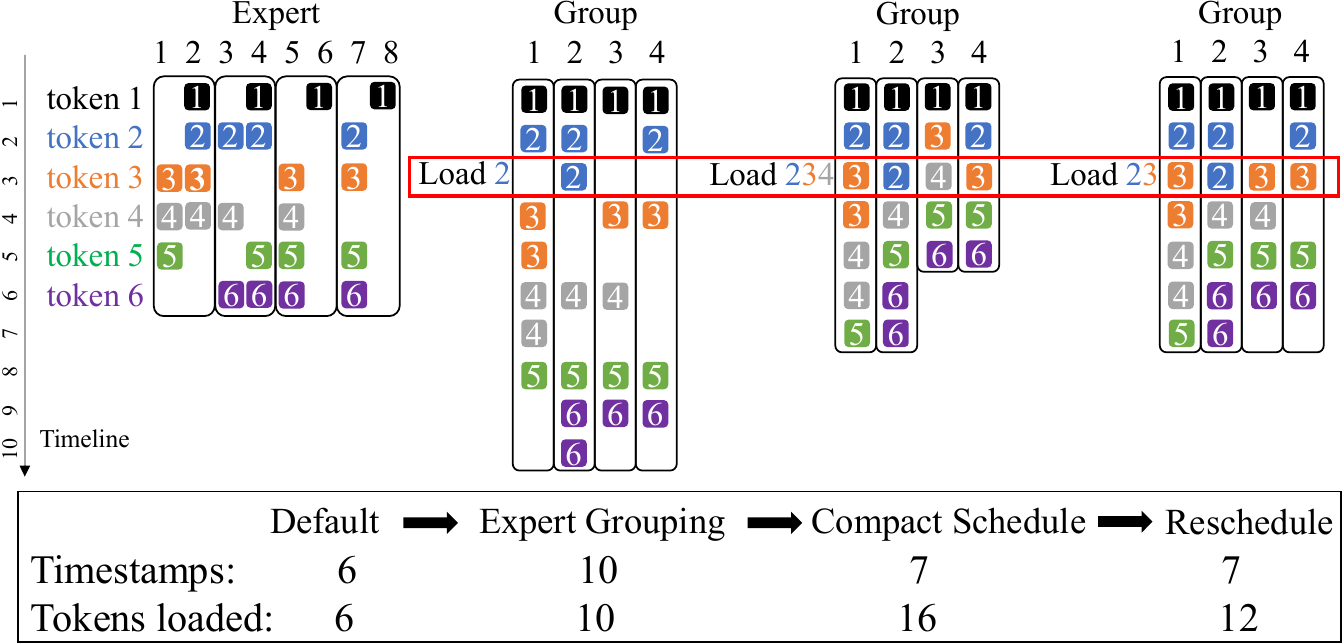}
\caption{Grouping and scheduling at the prefill stage.}
\label{fig:schedule}
\end{figure}

\begin{figure}[t]
\centering
\includegraphics[width=0.96\columnwidth]{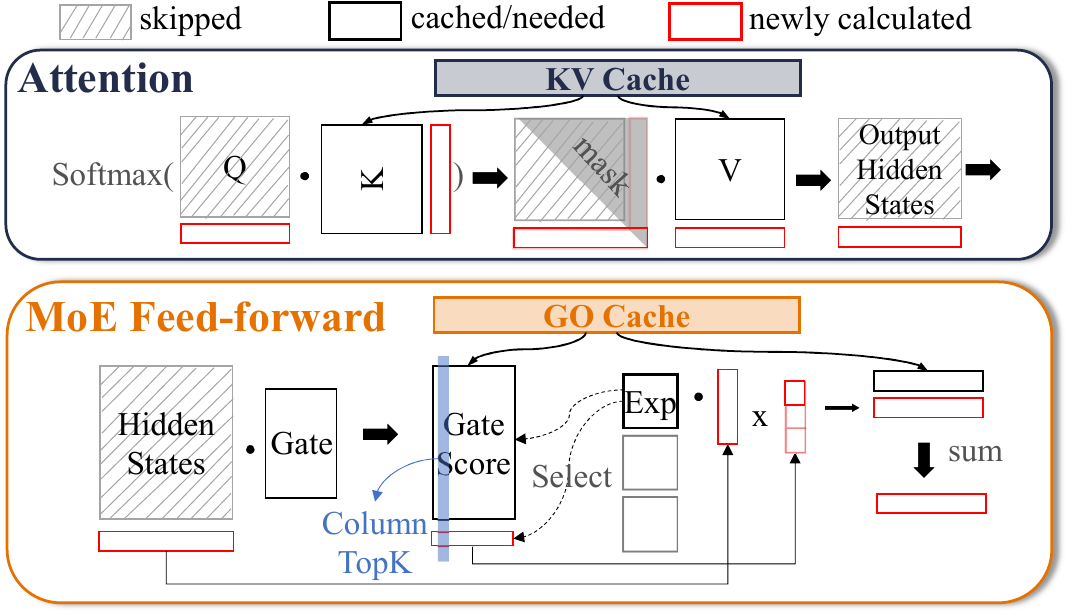}
\caption{The KV and GO cache for expert choice routing MoE. }
\label{fig:cache}
\end{figure}

\subsection{KVGO Cache for Expert Choice Routing}

To reduce redundant calculation during generation, we propose a gate-output (GO) cache for expert choice routing MoE, as illustrated in Fig.~\ref{fig:cache}.
We store the gate results and necessary outputs. The cache can be used together with the KV cache, and both are located in off-chip DRAM.
The future generation will bypass calculation and load the results from cache. Consequently, the gate receives only one token as the input during generation.
The Top-K results are calculated based on the previous scores and the incoming ones. If the new score is within the top-k results, the token will be selected. Gate caching is only required for expert choice routing, as the expert may change their top-k choices during generation. The gate computation in (\ref{eq:gate}) is modified as follows.

{\small
\setlength{\abovedisplayskip}{3pt} 
\setlength{\belowdisplayskip}{3pt}
\setlength{\abovedisplayshortskip}{2pt}
\setlength{\belowdisplayshortskip}{2pt}

\vspace{-5pt}
\begin{align}
    G(x) &= \mathrm{softmax}(\mathrm{TopKUpdate}(S_{\mathrm{prev}},\, x W_G ,\, k)) \label{eq:gate_cached}
\end{align}
\begin{equation}
\mathrm{TopKUpdate}(S_{\mathrm{prev}}, s, k)_j =
\begin{cases}
\begin{aligned}[t]
&\mathrm{TopK}\Big( S_{\mathrm{prev},j} \cup \{s_j\},\, k \Big),\\
&\text{\hspace{0.6cm} if } s_j \ge \min(S_{\mathrm{prev},j})
\end{aligned}
\\
S_{\mathrm{prev},j}, \quad \text{otherwise}
\end{cases}
\end{equation}
}

\noindent We use $S_{prev}$ to denote the scores of previous tokens from cache. After routing, only the experts who select the incoming token perform subsequent linear operations, consistent with (\ref{eq:calculator}).

If all tokens must be retained during generation, e.g., in constrained tasks \cite{post2018fast}, we further cache the final MoE output. We store the calculated top-k outputs of each expert. The storage will be $k\times \#experts \times d$, which is a static value and will not grow with token length.
We only cache $k$ results for each expert, so if the choices change, the cached contents change, but each generation step will result in at most one change per expert.
In addition, the cache size is much smaller than a normal linear layer due to the sparsity of the MoE. Consequently, in (\ref{eq:calculator}), $G(x)E(x)$ is retrieved directly from cache.

\begin{algorithm}[t]
\small
\caption{Reschedule by Inserting Idle}
\label{alg:schedule}
\begin{algorithmic}[1]
\Require $choices[T, E]$: token-to-expert choices; 

\State Group the choices and compute  $load[i, t]$ per group from the input
\State Identify the longest group: $max\_id$, length $L^*$

\State $csum[i,t] \gets$ cumulative sum of $load[i,1..t]$
\State $res[i,t] \gets csum[max\_id,t] - csum[i,t]$

\For{each group $i$} \Comment{in parallel}
    \For{each time $t$ where $res[i,t] > 0$} \Comment{pipeline}
        \State Insert $res[i,t]$ idles
               before the first element with data reuse (more positive numbers than others in $res[i, t..]$)
    \EndFor
    \State $res[i, t..] \gets res[i, t..] - res[i,t]$ 
    \State $Out[i] \gets choices_i (after \ grouping \ and \ insertion)$
\EndFor

\State \Return $Out$
\end{algorithmic}
\end{algorithm}

\subsection{Dynamic Scheduling Algorithm}

Conventional MoE inference follows a token-wise schedule, where tokens are fed to hardware one by one. However, this schedule has low utilization since different groups require different inputs.
In addition, token-expert choices are calculated on the fly.
Our workload-sorted grouping method considers overall load balancing statically. However, in the prefilling stage of an LLM, the gate network does not calculate the top-k results until tokens are fed to the MoE layer, so that static expert grouping cannot help its balance.
Therefore, we propose a dynamic scheduling method that reorders token processing within each group. The optimization opportunity comes from data reuse. Certain tokens may transfer repeatedly, especially considering the structural competition. The scheduling method finds the data transfer order that results in the least repeated data transfer. The schedule is only applied to the prefilling stage because, after the cache design, the generation process requires one token as input, regardless of token choice or expert choice.

The algorithm consists of two steps, as shown in Fig.~\ref{fig:schedule}.
First, instead of feeding tokens strictly one by one, we dispatch multiple tokens to different groups simultaneously to form a compact schedule. Second, we reschedule the order by inserting idle slots to avoid redundant data transfer. The insertion positions are determined by iteratively checking whether there is a data reuse opportunity. The example in Fig.~\ref{fig:schedule} shows that the number of transfers was reduced from 16 to 12. The detailed procedure is presented in Algorithm~\ref{alg:schedule}. The scheduling algorithm runs in linear time with respect to token length and can be pipelined in hardware, allowing its latency to be fully hidden.

\section{Experiments}\label{sec:experiment}




\subsection{Experimental Setup}

Our target model is Llama-MoE-4/16 \cite{zhu2024llama}, an MoE variant of Llama2-7B with 32 transformer blocks. The original MoE employs top-k token-choice routing; we implement expert-choice routing following \cite{zhou2022expertchoicerouting}, while keeping the model structure unchanged, thus avoiding performance degradation.
We simulate a single layer since all blocks have the same size.
Workload traces are sampled from the Pajama C4 dataset \cite{weber2024redpajama}. 

We evaluate PIM execution using an operator-level simulator built on 3DCIM \cite{buchel2025efficient}, where we implement the KVGO cache and our scheduling methods.
The PIM chip specification is HERMES \cite{khaddam2021hermes, khaddam2022hermes, le202364} (256 x 256 crossbar, 8-bit I/O).
The latency and power of activating one core are 130 $ns$ and 0.096 $nW$. The area is 0.635 $mm^2$.
Our model requires 1536 crossbars for 16 experts for one layer.
For all other components (operators, cache, DRAM, and digital units), we adopt the same assumptions or fit with polynomial functions as in \cite{buchel2025efficient}.
The GO cache is stored alongside the KV cache in off-chip DRAM
Each newly generated token only adds 32B of score data, and the output cache size is fixed at 512 KB.
The latency and power of the scheduler are constant with respect to token length and are therefore omitted.
For area evaluation and comparison, we report only the MoE linear cores, excluding off-chip DRAM and the digital part. 
In addition, to make a fair comparison in terms of area, we adopt the 2D manner layout for both our design and the baseline \cite{buchel2025efficient}.

In experiments, we use 32 prompt tokens and generate 8 to 64 new tokens.
The baseline is a direct deployment of 3DCIM without sharing, grouping, or scheduling, which means each crossbar exclusively occupies corresponding peripherals, and tokens are processed one by one. We evaluate group sizes of two and four. The uniform grouping and load-sorted grouping are denoted as $U$ and $S$, respectively.
For scheduling, the compact schedule and the optimized reschedule are denoted as $C$ and $O$.


\subsection{The Influence of Cache and Schedule}

\begin{figure}[t]
\centering
\begin{subfigure}[t]{0.48\columnwidth}
    \centering
    \includegraphics[width=\linewidth, trim=0 -25 0 0, clip]{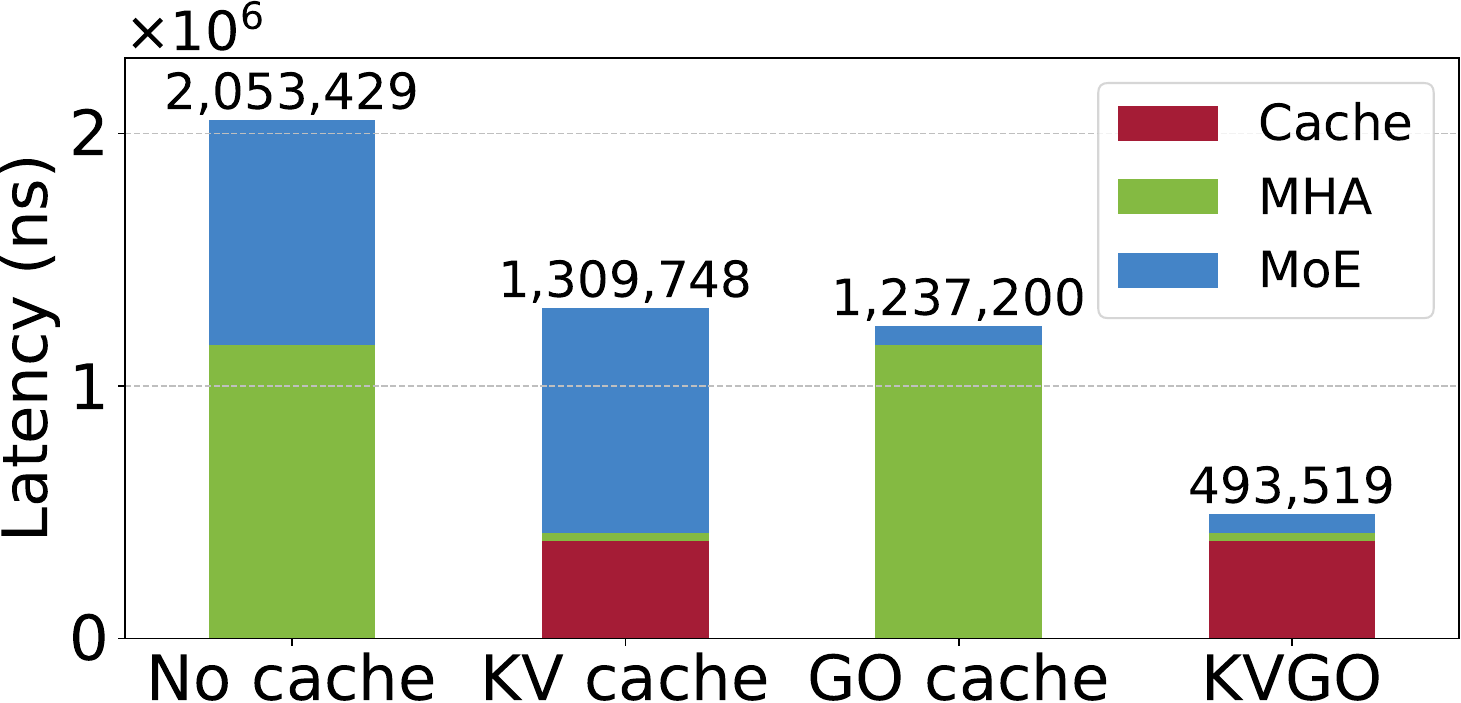}
    \caption{Latency of generating 8 tokens}
\end{subfigure}
\begin{subfigure}[t]{0.46\columnwidth}
    \centering
    \includegraphics[width=\linewidth]{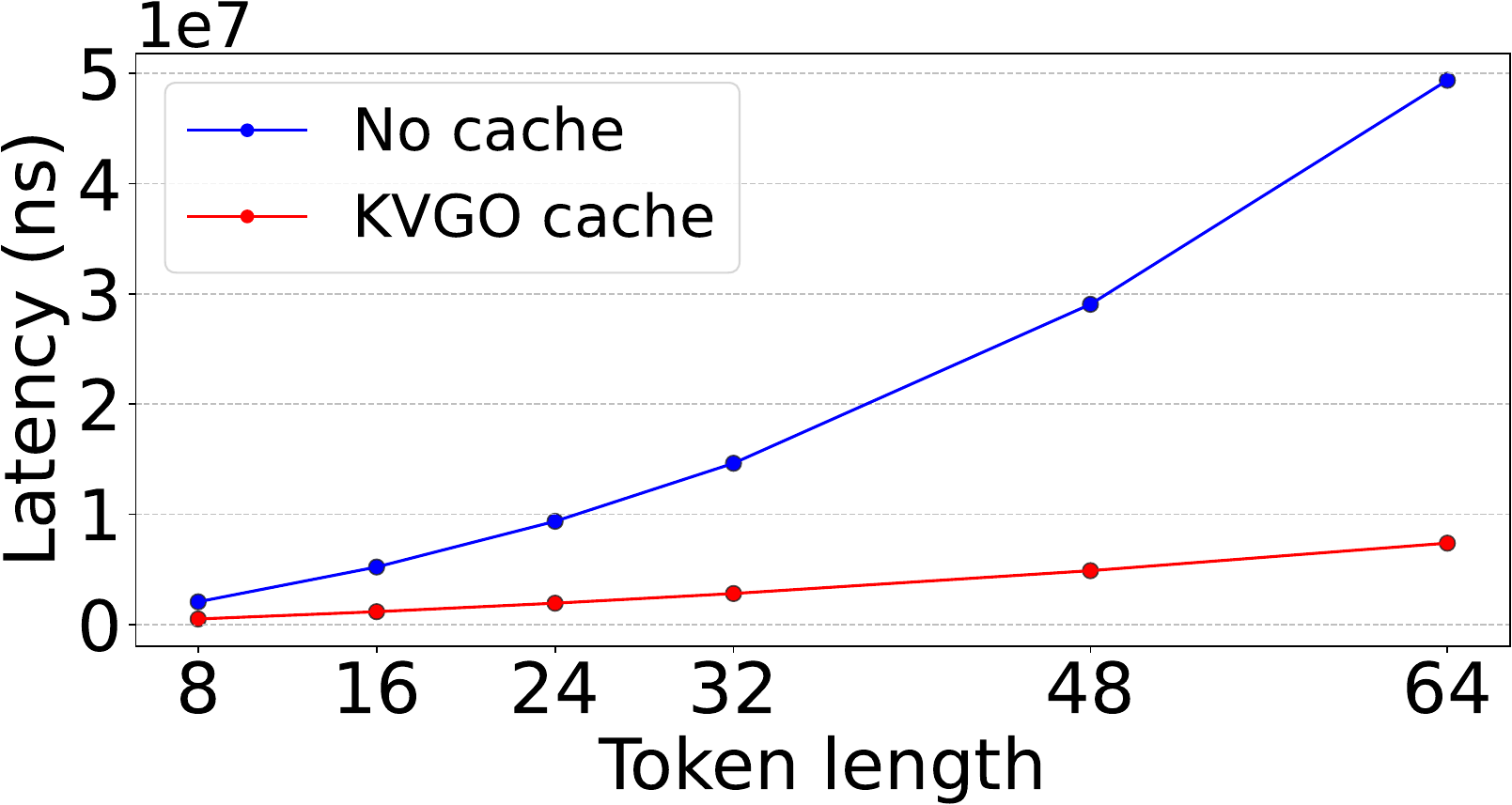}
    \caption{Latency varies with length}
\end{subfigure}
\caption{Latency results of the generate stage.}
\label{fig:cache_results}
\end{figure}

As shown in Fig.~\ref{fig:cache_results}, the KV cache reduces attention latency but does not benefit from energy because DRAM costs extra energy to transfer data.
The GO cache greatly reduces linear latency and energy because it allows execution with one token rather than all tokens as input during generation. 
According to Amdahl's law, improving a small part of the overall process will quickly lead to bottlenecks.
Therefore, our experiment demonstrates the necessity of the proposed GO cache in the generation task.
The maximized benefits come from the combination of the KV cache and the GO cache.
The latency and energy generating 8 tokens improve by 4.2x and 10.1x, respectively. Compared to KV cache, KVGO cache improves by 2.7x and 10.1x, respectively. The improvement will be larger when generating longer tokens. When generating 64 tokens, the latency improves by 6.7x, and the energy consumption improves by 14.1x.
The result of latency with token length is shown in Fig.~\ref{fig:cache_results} (b). Our cache design has better scalability and grows linearly with the token length.

\begin{figure}[t]
\centering
\begin{subfigure}[t]{0.32\columnwidth}
    \centering
    \includegraphics[width=\linewidth]{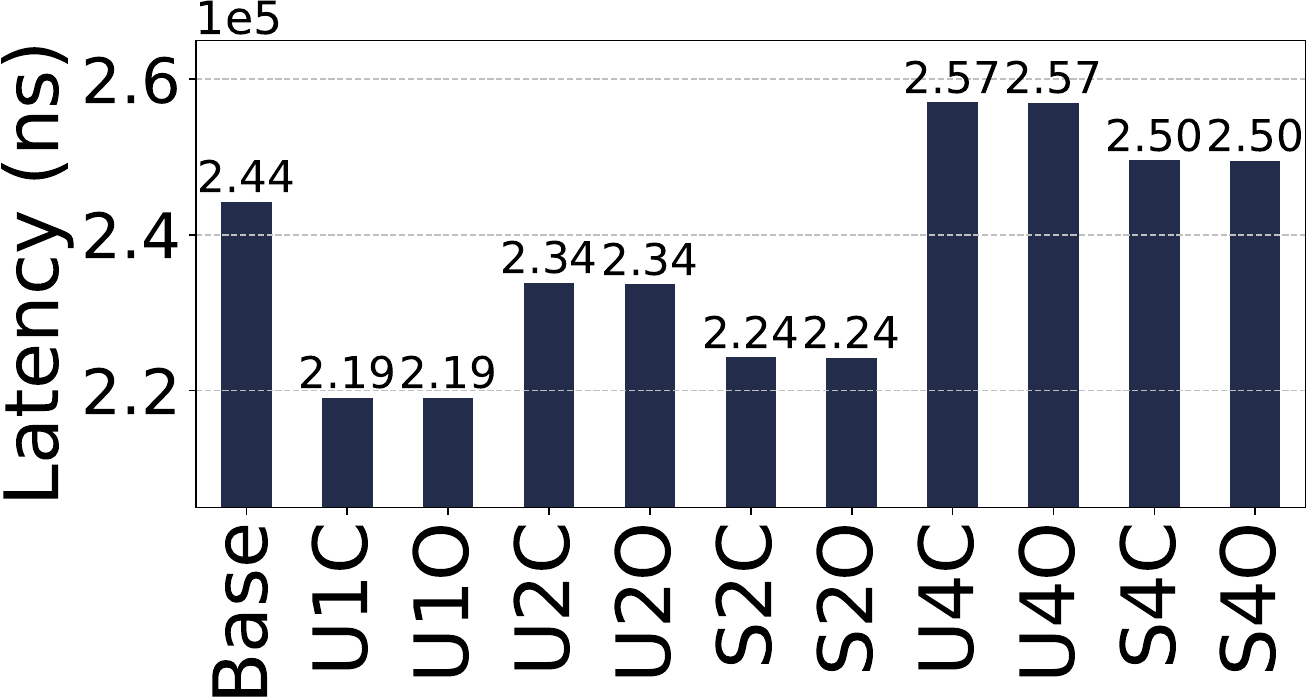}
    \caption{Latency}
\end{subfigure}
\begin{subfigure}[t]{0.32\columnwidth}
    \centering
    \includegraphics[width=\linewidth]{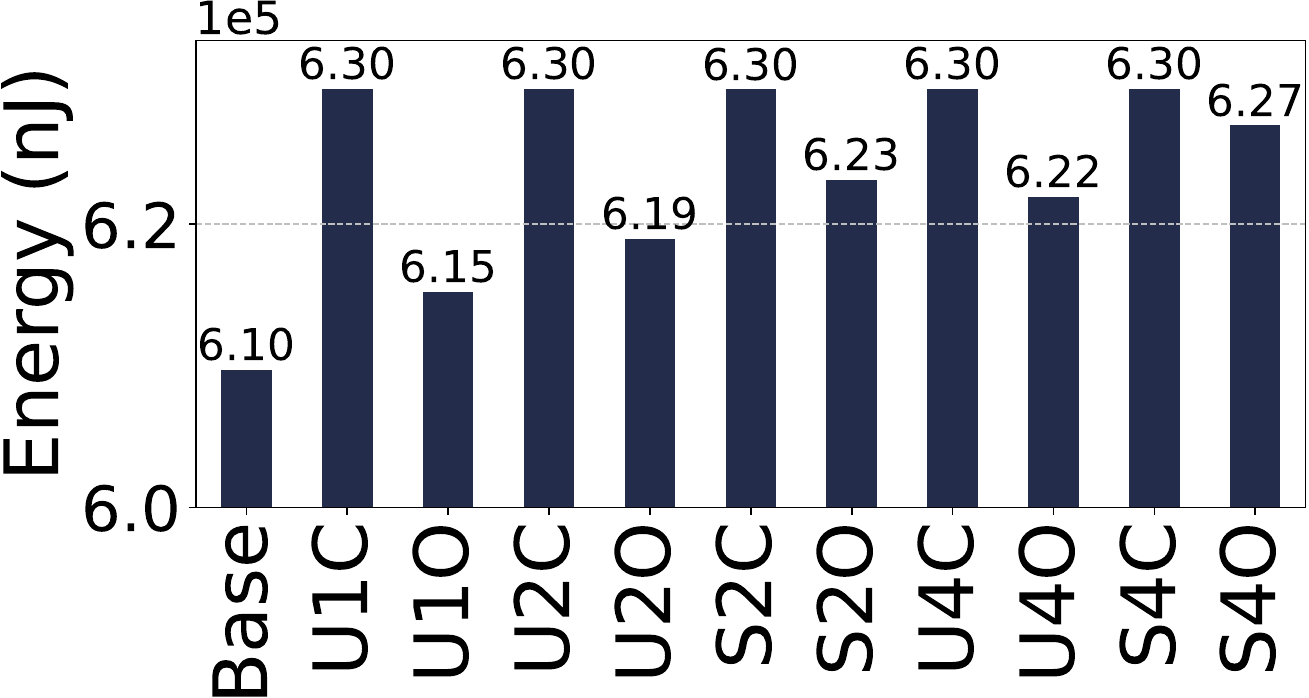}
    \caption{Energy}
\end{subfigure}
\begin{subfigure}[t]{0.30\columnwidth}
    \centering
    \includegraphics[width=\linewidth]{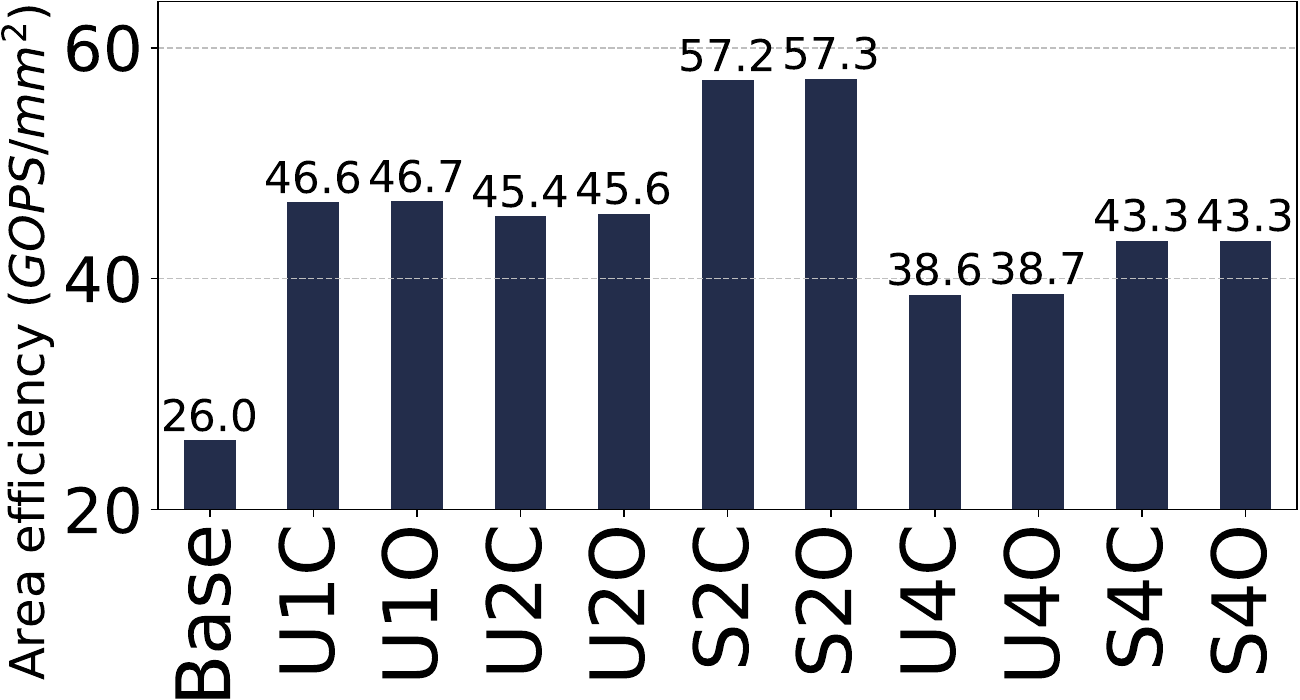}
    \caption{Area efficiency}
\end{subfigure}
\caption{Results of different scheduling methods. }
\label{fig:schedule_results}
\end{figure}

The grouping and scheduling results are shown in Fig.~\ref{fig:schedule_results}. The load-sorted grouping improves the latency result due to the balanced load. In addition, a large group reduces the total area but incurs more competition, which hinders performance. In our experiments, a group of two experts gained the best area efficiency ($GOPS/mm^2$) result.
Following our experimental setup, the crossbar area accounts for 40\% of the total area, making the benefits of a larger group relatively low. The potential gain can be increased if we generalize the case with a smaller crossbar area ratio. For example, with \cite{shafiee2016isaac}, we can gain more benefits with a large group size, i.e., 4, where our design reaches 82.7 $GOPS/mm^2$ under a crossbar area ratio of 5\%.
From Fig.~\ref{fig:schedule_results}, the compact schedule reduces the latency but causes repeated data transfer and a slight increase in energy consumption. Our reschedule algorithm improves both latency and energy because it has the latency of a compact schedule and less repeated data transfer. When counting area, all scheduling methods are beneficial.
Our approach (S2O) improves efficiency by up to 2.2x over the baseline.

\subsection{Performance, Energy, and Density Results}

\begin{table}[ht]
\caption{Total Latency, Energy, and Density Comparison}
\begin{center}
\label{tab:compare}
\setlength{\tabcolsep}{6pt}
\renewcommand{\arraystretch}{1.2}
\begin{adjustbox}{width=\columnwidth}
\begin{tabular}{lccc}
\hline
 & \textbf{No cache, No schedule} & \textbf{KVGO cache, S2O} & \textbf{KVGO cache, S4O} \\
\hline
Latency ($ns$) & 2,297,724 & 717,752  & 743,078 \\
Energy ($nJ$) & 5,393,776 & 1,096,691 & 1,100,548 \\
\hline
\makecell{Performance Density \\ ($GOPS/W/mm^2$)} & 10.2 & 12.3 & 15.6 \\
\hline
\end{tabular}
\end{adjustbox}
\end{center}
\end{table}
The total latency, energy, and efficiency results, including prefill and generate stages, are shown in TABLE~\ref{tab:compare}. 
The best performance and energy of a complete inference come from S2O with KVGO cache, which improves latency and energy by 3.20x and 4.92x.
The S4O with KVGO cache reaches the best performance density ($GOPS/W/mm^2$) and improves by 1.53x compared to the baseline.
The overall improvement benefits from scheduling from the prefill stage and cache design for accelerating generation.

\section{Conclusion}

In this paper, we propose an area-efficient PIM accelerator for MoE by multiplexing, scheduling, and GO cache. Future works focus on hardware-aware software design and noise analysis.




\bibliographystyle{ieeetr}
\bibliography{reference}

\end{document}